\documentclass{article}
\usepackage[onecolumn]{emulateapj}

\def\brfrac#1#2{\left(\dfrac{#1}{#2}\right)}
\def\dfrac#1#2{{\displaystyle\frac{\mathstrut #1}{#2}}}
\def\pa#1#2{\biggl( \dfrac{\partial #1}{\partial #2} \biggr)}
\def\ea{{\rm et~al.\ }}

\def\ltsima{$\; \buildrel < \over \sim \;$}
\def\lsim{\lower.5ex\hbox{\ltsima}}
\def\rmAA{{\rm \AA}}
\def\Mdot{{\dot{M}}}
\def\Mbh{{M_{\rm BH}}}
\def\Msun{{M_\odot}}
\def\Ledd{{L_{\rm Edd}}}
\def\Fnu{{F_\nu}}
\def\Rg{{R_{\rm g}}}
\def\t0{{\tau_0}}
\def\tc{{\tau_{\rm c}}}
\def\sigmaT{{\sigma_{\rm T}}}
\def\sigmaffnu{{\sigma_{{\rm ff}, \nu}}}
\def\jffnu{{j_{{\rm ff}, \nu}}}
\def\jcompnu{{j_{{\rm Comp}, \nu}}}
\def\lambdaie{{\lambda_{\rm ie}}}
\def\OmegaK{{\Omega_{\rm K}}}
\def\qrad{{q_{\rm rad}^-}}

\def\Qadv{{Q^-_{\rm adv}}}

\lefthead{Kawaguchi, Shimura and Mineshige}
\righthead{Broad-Band SEDs of AGNs from an Accretion Disk with Advective 
Coronal Flow}

\begin{document}
\submitted{Accepted for publication in the Astrophysical Journal}
\title{BROAD-BAND SPECTRAL ENERGY DISTRIBUTIONS OF ACTIVE GALACTIC NUCLEI 
FROM  AN ACCRETION DISK WITH ADVECTIVE CORONAL FLOW}
\author{Toshihiro Kawaguchi$^{1,2}$, Toshiya Shimura$^{3}$, 
{\rm and} Shin Mineshige$^{1}$}
\affil
{$^1$:  Department of Astronomy, Graduate School of Science, 
Kyoto University, Sakyo-ku, Kyoto 606-8502, Japan}
\affil
{$^2$:  Research Fellow of the Japan Society for the Promotion of
Science}
\affil
{$^3$:  International Graduate School of Social Sciences, 
Yokohama National University, Hodogaya-ku, Yokohama, 240-8501, Japan}
\centerline{e-mail: kawaguti@kusastro.kyoto-u.ac.jp}
\authoremail{kawaguti@kusastro.kyoto-u.ac.jp}
\begin{abstract}
Recent multi-waveband observations of Seyfert nuclei and QSOs
 established significant deviations in the spectral shape of the 
 big blue bump from a blackbody one;
Soft X-ray excess has a spectral index $\alpha$ 
($\Fnu \propto \nu^{- \alpha}$)
of 1.6 and hard X-ray tail with $\alpha$ of $\sim$ 0.7.
We construct a disk-corona model which accounts
for such broad-band spectral properties. 
We study emission spectrum emerging from a vertical 
disk-corona structure composed of two-temperature plasma by solving 
hydrostatic equilibrium and radiative transfer self-consistently.
A fraction $f$ of viscous heating due to mass accretion is assumed to be
dissipated in a corona with a Thomson optical depth of $\tc$, where
advective cooling is also included, and a remaining fraction, $1-f$,
dissipates within a main body of the disk. 
Our model can nicely reproduce the soft X-ray excess 
with a power-law shape and the hard tail 
extending to $\sim$ 50 keV.
The different spectral slopes ($\alpha \sim$ 1.5 below 2keV and $\sim$ 0.5
above) are the results of different emission mechanisms and different sites;
the former slope is due to 
unsaturated Comptonization from the innermost zone
and the latter is due to a combination of the Comptonization,
bremsstrahlung 
and a reflection of the coronal radiation at the disk-corona boundary
from the inner to surrounding zone ($\leq$ 300 Schwarzschild radii).
The emergent optical spectrum is redder 
($\alpha \sim 0.3$) than that of the standard disk 
($\alpha \sim -0.3$), being consistent with observations,
due to the different efficiencies of spectral distortion of disk 
emission at different radii. 
Further, we find that the cut-off frequency of the hard X-ray ($\sim$
 coronal electron temperature) and broad-band spectral shape 
are insensitive to the black-hole mass, 
while the peak frequency of the big blue bump 
is sensitive to the mass as the peak frequency $\propto \Mbh^{-1/4}$.
\end{abstract}

\vspace{5mm}

\section{INTRODUCTION}

Active Galactic Nuclei (AGNs) exhibit a lot of spectral components 
over a wide waveband; 
optical/ultraviolet (UV) bump, power-law component in hard X-rays, 
excess component in soft X-ray band (so called soft excess), 
warm absorber feature around 1 keV in some objects, 
and reflection/Fe K${\alpha}$ 
fluorescence line around a few to a few tens of keV 
(Mushotzky, Done \& Pounds 1993; Koratkar \& Blaes 1999 for reviews). 
Those spectral components are commonly thought to be powered 
by gas accretion onto a massive black hole. 
The most well-known disk-accretion model is called as the standard disk
model (Shakura \& Sunyaev 1973).
According to the standard disk model, the spectrum at 
each radius of the disk is assumed to be a blackbody radiation with 
a local effective temperature, $T_{\rm eff}$.
This simple picture was supported by their luminosity and 
by rough agreement of 
optical/UV spectral energy distribution (SED) between 
observations and models (e.g., Shields 1978; Malkan \& Sargent 1982).

The standard model, however, has limitations;
(i) if UV turn-over 
around 2000 $\rmAA$ is indeed an indication of the temperature 
at the innermost radius (e.g., Malkan 1983; Sun \& Malkan 1989) 
such disks are too cool to produce enough soft X-ray photons. 
(ii) Inversely, if the soft excess component at 0.1 to 1 keV is due to 
radiation from the innermost region, as is often interpreted, such
a disk only produces by 1.5--2 orders of magnitude
less optical/UV flux than what is observed.
(iii) When optical and soft X-rays spectra are simultaneously fitted 
with the disk spectrum, 
the luminosity of the disks ($L$) often exceeds the Eddington
luminosity ($\Ledd$).
(iv) Observed optical spectra of QSOs are typically redder 
($\alpha \sim 0.3$, where $L_\nu \propto \nu^{-\alpha}$; 
Francis et al.\ 1991) than those of the 
simplest standard accretion disks ($\alpha \sim - 1/3$).
In other words, a successful model spectrum of AGNs should deviate from 
that of the standard accretion disk.
(v) Hard power-law X-ray cannot be reproduced.

Then, a number of authors have tried to distort the disk spectrum toward 
the high energy regime so that the disk can emit substantial soft 
X-ray and optical/UV radiation simultaneously.
One promising idea is the 
Comptonization within the disk in the vertical direction 
(Czerny \& Elvis 1987; Wandel \& Petrosian 1988; Laor \& Netzer 1989; 
Ross, Fabian \& Mineshige 1992). 
The most accurate treatment of Comptonization 
in the framework of the standard model was made by 
Shimura \& Takahara (1993, hereafter ST93) who solved radiative transfer 
and vertical structure simultaneously and presented 
emergent spectra integrated over radii 
(see also Shimura \& Takahara 1995, hereafter ST95).
The effect of Comptonization is more prominent at higher accretion rates
(e.g., Ross et al.\ 1992; ST93; ST95). 

However, there still remain discrepancies between
models and observations. 
(i) Although Comptonization tends to increase $\alpha$, 
the Far-UV (FUV) spectra of 
these accretion disks exhibits $\alpha$ $\sim$ 1 at best
when the luminosity of the disk approaches the 
Eddington limit (Ross et al.\ 1992; ST95).
On the other hand, the observations of distant quasars showed 
steeper FUV spectra ($\alpha \sim$ 1.8 -- 2.2, Zheng et al.\ 1997). 
(ii) The observed small spectral indices in soft X-ray 
($\alpha \sim 1.4$ -- $1.6$; e.g., Walter \& Fink 1993; Laor et
al. 1997) are not achieved by any disk models 
(e.g., Nandra et al.\ 1995 for Mrk 841; 
Laor \ea 1997 for a sample of low-redshift QSOs), 
because the higher energy tail of those disk models 
is the superposition of Wien laws, thus exhibiting exponential roll-over. 
(iii) These accretion disks still cannot reproduce the hard
X-rays. 
Then, hard X-ray emission should be treated as additional components 
in these models.

With these problems kept in mind, we, in this present study, aim 
to produce the overall SED simultaneously by disk-corona models.
The spectrum to be reproduced with these models is a composite
one obtained from several independent observations 
(see Zheng et al.\ 1997; Laor et al.\ 1997). 
It is composed of `typical' spectral indices over broad-bands from 
Near-IR to hard X-ray, optical--to--X-ray flux ratio,
and the energy cut-off of the
hard power-law component (see \S 3.1 for more detailed description). 
We can, for the first time, reproduce such broad-band spectra.
In \S 2 we review the assumptions and numerical method used in the
calculation. 
Numerical results and comparisons between the models 
and the composite spectrum are presented in \S 3. 
The final section is devoted to discussion and summary.

\section{BASIC ASSUMPTIONS AND EQUATIONS}

The numerical code used in this study is basically the same as that of
ST93 except for some modifications.
We assume for the configuration of the system that
the accretion disk main body is sandwiched between coronal layers in the
vertical direction (e.g., Haardt \& Maraschi 1991), 
and that the whole system is geometrically thin 
(i.e., plane-parallel slab geometry).
We treat the disk-corona system consisting of fully ionized hydrogen,
thermal plasma around a Schwarzschild black hole of mass $\Mbh$.
Gas evaporation from the disk to corona and -condensation from corona to disk 
(e.g., Meyer \& Meyer-Hofmeister 1994) are not included here for simplicity;
the disk and coronal layers are interacting only via radiation and
pressure (e.g., Nakamura \& Osaki 1993; 
\.{Z}ycki, Collin-Souffrin \& Czerny 1995).
Although magnetic fields may also affect the disk-corona structures,
they are not included in this calculations.

The equation of hydrostatic equilibrium in the vertical direction 
is given by
\begin{eqnarray}
 -\frac{d P_{\rm gas}}{dz} + \int^{\infty}_{0} 
  d\nu \frac{\Fnu}{\lambda_{\nu}c} 
  = m_{\rm p} N_{\rm e} \frac{G \Mbh}{R^3} z,
\end{eqnarray}
where $R$ and $z$ are the radial and the vertical coordinates, respectively;
and $P_{\rm gas}$ and $N_{\rm e}$ are the gas pressure and the number 
density of electrons, respectively.
For the radiation field, we adopt the diffusion approximation.
Then, the radiative flux $\Fnu$ at some frequency $\nu$ is given by
\begin{eqnarray}
 \Fnu = - \frac{c \lambda_\nu}{3} \pa{\epsilon_\nu}{z},
\end{eqnarray}
where $\epsilon_\nu$ and $\lambda_\nu$ are the radiation-energy density
per unit frequency per unit volume and the mean free path of a photon 
with a frequency $\nu$, respectively. 
We set $\lambda_\nu = (\sigmaT + \sigmaffnu)^{-1} N_e^{-1}$;
we do not include the bound-free nor free-bound transitions 
in the calculations.
Here, $\sigmaT$ and $\sigmaffnu$ are the cross sections of Thomson
scattering and bremsstrahlung absorption, respectively.
The Thomson optical depth of the disk-corona measured from the 
mid-plane $\tau$ is related to the height from the mid-plane $z$, as
\begin{eqnarray}
 d \tau = N_e \, \sigmaT \, dz.
\end{eqnarray}
The total Thomson optical depth from the mid-plane (through the boundary
between the disk-corona) to the surface of the corona, $\t0$, 
is a free parameter.
The equation of state is that $P_{\rm gas} = N_e k_{\rm B} (T_p + T_e)$, where 
$T_p$ and $T_e$ are temperature of protons and electrons, respectively.

The dissipated energy per unit surface area of the disk-corona system
$Q^+_0$ in Newtonian approximation is written as (Shakura \& Sunyaev 1973)
\begin{eqnarray}
 Q^+_0 = \frac{3}{8 \pi} \Mdot \OmegaK^2 \left(1-\sqrt{3 \Rg / R} \right),
\end{eqnarray}
where $\Mdot$ and $\OmegaK$ [$=(G \Mbh / R^3)^{1/2}$] are the total 
(above and below the mid-plane) 
accretion rate and the Keplerian angular velocity, respectively. 
The innermost radius of the disk-corona is assumed to be $3 \Rg$, where 
$\Rg$ [$= 2 G \Mbh / c^2 = 3 \times 10^{13} {\rm cm} (\Mbh / 10^8 \Msun)) = 
0.01 \!$ lt-day $\!(\Mbh / 10^8 \Msun)$] 
is the Schwarzschild radius.
For the case of a non-rotating black hole under non-relativistic treatment, 
$\Mdot = 12 \Ledd / c^2$ [$= 2.6 \, \Msun {\rm yr}^{-1} 
(\Mbh / 10^8 \Msun)$] corresponds to the accretion disk shining at 
$\Ledd$ [$= 1.3 \times 10^{46} {\rm erg} \ {\rm sec}^{-1} 
(\Mbh / 10^8 \Msun)$]. 

A constant fraction $f$ of mass accretion is assumed to be
dissipated in the corona with a Thomson optical depth of $\tc$, where
advective energy transport of protons is also included 
in addition to radiative cooling of electrons (see Figure~1). 
A remaining fraction, $1-f$, dissipates within the disk layer. 
Advection in the disk layer (i.e., slim disk model that will be
discussed later) is not included because of
small radial velocity there.
The advective cooling rate in the corona per unit surface area, 
$\Qadv$, is taken from the expression of optically-thin advection
dominated accretion flow (ADAF) in one-temperature case 
(i.e., $T_{\rm p} = T_{\rm e}$; see, Kato, Fukue \& Mineshige 1998, p272):
\begin{eqnarray}
 \Qadv = \frac{1}{16 \pi^2} \brfrac{\OmegaK}{R^2} 
  \frac{(f \Mdot)^2 \sigmaT}{\alpha_c m_p \tau_c}.
\end{eqnarray} 
Here, $\alpha_c$ is the viscosity parameter which controls the efficiency 
of the advection in the corona.
We perform numerical calculations under the condition that 
the advective cooling should be less
than the dissipated energy in the coronal layer; 
$\Qadv \leq f \ Q^+_0$.

The heating rate in the disk and corona, 
$q^+_{\rm d}$ and $q^+_{\rm c}$, respectively, 
and advective cooling rate in the corona per unit volume, 
$q^-_{\rm adv}$, are assumed to be 
proportional to matter density at each site.
As a result, the fraction of advective cooling in the dissipated energy
is 
\begin{eqnarray}
 \frac{q^-_{\rm adv}}{q^+_{\rm c}} = \frac{\Qadv}{f \, Q^+_0} =
  0.21 \frac{f}{\alpha_c \tc} \brfrac{\Mdot}{\Ledd / c^2} 
  \brfrac{R}{5 \Rg}^{-1/2}
  \left(1 - \sqrt{\frac{3 \Rg}{R}} \right)^{-1}. 
\end{eqnarray}
The energy balance in each layer is as follows:
\begin{eqnarray}
{\rm Disk}  : &{\rm Protons  }& : \quad q^+_{\rm d} = \lambdaie \\
               &{\rm Electrons}& : \quad \lambdaie = \qrad \\
{\rm Corona}: &{\rm Protons  }& : \quad q^+_{\rm c} = q^-_{\rm adv} + \lambdaie \\
               &{\rm Electrons}& : \quad \lambdaie = \qrad,
\end{eqnarray}
with
\begin{eqnarray}
 q^+_{\rm d} = (1-f) Q^+_0 \frac{\sigmaT N_e}{\t0 - \tc},  \qquad
 q^+_{\rm c} =   f   Q^+_0 \frac{\sigmaT N_e}{\tc},  \qquad
 q^-_{\rm adv}= \Qadv \frac{\sigmaT N_e}{\tc},
\end{eqnarray}
where $\lambdaie$ is the energy exchange rate due to 
Coulomb collisions taken from Guilbert and Stepney (1985); 
and $\qrad$ is the radiative cooling rate.
We consider bremsstrahlung and Comptonization for emission mechanisms.
Then, radiative cooling rate $\qrad$ is described as 
\begin{eqnarray}
 \qrad = \int d\nu (-c N_e \sigmaffnu \epsilon_\nu + \jffnu + \jcompnu),
\end{eqnarray}
where $\jffnu$ and $\jcompnu$ are the bremsstrahlung emissivity 
(Rybicki \& Lightman 1979) 
and the net rate of energy transfer 
from electrons to photons via Comptonization 
per unit volume per unit frequency, respectively. 
The latter is described by the Kompaneets equation 
(Rybicki \& Lightman 1979; see Hua \& Titarchuk 1995 for comparisons of 
the analytical treatment of Comptonization with Monte Carlo simulations).
Thus, the radiative transfer equation for $\Fnu$ is written as 
\begin{eqnarray}
 \frac{\partial \Fnu}{\partial z} = 
  -c N_e \sigmaffnu \epsilon_\nu + \jffnu + \jcompnu.
\end{eqnarray}
For the expression of the free-free absorption and emission
($\sigmaffnu$ and $\jffnu$, respectively), 
we take the Gaunt factor to be unity, for simplicity (see
Rybicki \& Lightman 1979).

Following ST93, we 
take $\xi$ [\,$\equiv$log($\t0 - \tau$)\,] as an independent variable 
for the vertical coordinate. 
To calculate the spectra integrated over the whole disk (\S 3.1), 
we divide the disk-corona from $3 \Rg$ to $300 \Rg$ into 20 consentric rings 
so that each ring radiates approximately 
the same luminosity (cf. Ross et al.\ 1992; ST95).
In total, input free parameters required for the calculations are
$\Mbh$, $\Mdot$, $f$, $\tc$, $\t0$, and $\alpha_c$. 
The number of these 
parameters is similar to that of relevant observed parameters 
which we aim to explain simultaneously; 
e.g., $L_X$, $\alpha_{\rm ox}$, $\alpha_{\rm UV}$, 
$\alpha_{\rm opt}$, $\alpha_{ROSAT}$, $\alpha_{ASCA}$, etc. 
Then, outputs are 
$N_e(\xi)$, $T_e(\xi)$, $T_i(\xi)$, $P_{\rm gas} (\xi)$, $z(\xi)$, and 
$\epsilon_\nu (\xi)$.
Spectrum of the whole disk-corona system is obtained by summing up the
emergent spectra of all the rings.

The emergent spectrum at each ring does not so strongly depend on 
$\t0$ as long as $\t0 \gg 1$, 
but is sensitive to $\tc$.
We, thus, take $\t0$ as a constant over all rings, for simplicity.
Currently, we do not have a good theory to predict 
the radial dependence of $\tc$ and $f$ that can, 
in principle, be determined by physics of evaporation/condensation.
Thus, $\tc$ and $f$ are also simply assumed to be constant over all rings.
The effects of changing  $\tc$ and $f$ will be discussed later.

General relativistic effect is not included.
Since no Doppler broadening is considered, the total spectrum represents 
the case of a face-on disk. 
Convection/conductive energy transport is not included. 
Shakura, Sunyaev \& Zilitinkevich (1978) have shown that convection
transports no more than $\sim$ 30 \% of the vertical energy flux.

\section{SPECTRAL ENERGY DISTRIBUTION OF AGNS}

\subsection{Observed Composite Spectrum}

The spectrum to be reproduced with our models is a composite
one (dotted lines in Figure 2a), 
which is a useful probe to see whether a model for the SED of
AGNs can work or not. 
The vertical normalization is determined so as to give rise to 
a representative optical luminosity among 
low-redshift quasars from the Bright Quasar Survey adopting $H_0$ of 50 km
s$^{-1}$ Mpc$^{-1}$ and $q_0$ of 0.5 with an assumption of isotropic
emission (see Laor et al.\ 1997).
There are, however, two major problems that one should keep in mind
when dealing with the composite spectrum (Koratkar \& Blaes 1999; Laor 1999).
(i) We observe objects with different redshifts 
in different wavebands; spectral index in
FUV is obtained from distant QSOs (Zheng et al.\ 1997), 
while that of soft X-ray is mainly from nearby objects (e.g., Walter \&
Fink 1993; Laor et al.\ 1997). 
Also, sample used in each wavebands contains objects that 
are not necessarily the same objects.
(ii) Soft excess might be an instrumental/calibration problem; 
{\it BeppoSAX} observations did not 
detect soft X-ray excess in some objects 
while excess was seen for the same objects with 
other telescopes (e.g., Matt 1999). 
Although the results are still controversial,
we assume that soft excess really exists in most AGNs 
throughout this paper. 
In what follows, 
we try to reproduce broad-band spectra similar to the composite one, 
as the first step.

\subsection{Model Spectrum Integrated over Radii}

We first show the most successful case with $\Mbh =3 \times 10^9 \Msun$
and $\Mdot = 0.5 \ \Ledd / c^2$
that gives rise to luminosity of about 5 \% of the Eddington limit
for a non-rotating black hole under non-relativistic treatment.
Thick line in figure 2a shows an example of the resultant broad-band 
spectra.
Since the innermost ring from 3.0 $\Rg$ to 4.0 $\Rg$ does not satisfy
the restriction that $q^-_{\rm adv} / q^+_{\rm c} \leq 1.0$ 
for the parameter sets described in Fig.\ 2a, 
we, hereafter, plot integrated spectra from 4.0 $\Rg$ to 300 $\Rg$.
In other words, all dissipated energy is assumed to be carried out by
advection in $3.0 < R < 4.0 \Rg$. 
To account for the radiation at the surfaces of coronae 
above and beyond the disk,
the resultant spectrum is multiplied by two.
A significant fraction $f = 0.6$ of mass accretion occurs at the
corona.
In this case, advective cooling in the corona is comparable with the
radiative cooling at $R \sim 4.9 \Rg$; $q^-_{\rm adv} / q^+_{\rm c} \sim 0.5$ 
[eq.\ (6)].
A spectrum of the standard disk with the same $\Mbh$ and $\Mdot$ 
is also depicted for comparison (dashed line).

Presence of multiple spectral components is the most noteworthy feature
of the present model. 
This is because different radiative mechanisms play roles 
in different wavebands in Fig.\ 2a; 
thermal radiation of the disk in optical/UV,
unsaturated Comptonization in FUV/soft X-ray, and a combination of the
power-law component due to Comptonization, 
bremsstrahlung, and a reflection in hard X-ray.
Note that the underlying radiative processes in soft--hard X-rays are
distinct from those of the traditional explanation, in which UV--soft X-ray
component is due to blackbody whereas 
hard power-law component is due to Comptonization of the soft photons.
In our model, hard X-ray spectrum 
looks a power-law with $\alpha \sim$ 0.5--1.0 due to 
the combination of multiple radiative mechanisms, in contrast.

Figure 2b shows contributions to the total spectrum ($4.0 \Rg < R < 300
\Rg$; thick line) due to individual rings.
The outermost ring ($104 \Rg < R < 300 \Rg$) contribute as much as one
third of the total spectrum at a few tens of keV.
Usual models for X-ray emission of AGNs assume that 
only inner region radiates X-ray (i.e., one-zone);
our model differs from the traditional 
model in terms of radial dependence of the X-ray spctrum, as well.
Like optically-thin ADAF (e.g.,
Narayan, Yi \& Mahadevan 1995; Manmoto, Mineshige \& Kusunose, 1997), 
X-ray emission arises from wide spatial range ($R \lesssim 300 \Rg$), 
in contrast with a usual belief.
Interestingly, contribution from the outer rings to the total X-ray 
spectrum decreases towards lower-energy X-ray band. 
Provided that inner rings are more time-variable than outer rings, 
resultant X-ray spectrum will get softer when the luminosity increases,
as is actually observed (e.g., Done \ea 1995; Leighly \ea 1996).
The most powerful test of our model will be gravitational microlensing
(Yonehara et al. 1998) which provides information as to the size of
emission region as a function of wavelength on AU ($\sim \Rg$ for $10^8
\Mbh$) scales. 

The effect of advective cooling in the
coronal layer to the emergent spectrum is demonstrated in
Figure 2c. 
Thick dashed line is the resultant spectrum without advective cooling, 
which corresponds to a two-temperature treatment of 
the study by Shimura, Mineshige, \& Takahara (1995), but
with a constant $\tc$.
Lower dashed curves represents the spectra of each ring in the case
without advective cooling.
Thick solid curve has the same meaning as in Fig.\ 2a and 2b.

In our models above (Fig.\ 2a--2c), we assumed constant $f$ and $\tc$
over all rings.
This will be a reasonable assumption since 
Janiuk, \.{Z}ycki \& Czerny (2000), for example, estimated $f$ as a 
function of radius under some assumptions, finding that $f$ is a
slowly increasing variable with radius, and that 0.2 $< f <$ 1.0 for various
parameter sets at 3 to 300 $\Rg$.
The coronal gas is probably originated in evaporated gas cumulated in
the flow (Meyer \& Meyer-Hofmeister 1994; 
Liu, Meyer \& Meyer-Hofmeister 1997).
In other words, we have assumed that a disk corona has already 
formed until $\sim$ 300 $\Rg$.
If, however, evaporation of disk material is still active 
at $R \lesssim 300 \Rg$,
$\tc$ and $f$ are both likely to increase with a decreasing radius.
To what extent the resultant spectrum changes in such a case is
demonstarated in figure 2d.
Spectra for the outermost ring with $f = \tc = 0.6$ (solid curve) and 
$f = \tc = 0.1$ (dashed curve) are shown.
In the latter case, hard X-ray emission is not strong as the former
case, in which bremsstrahlung emission at the outermost ring is the main
origin of hardening in the total X-ray spectrum (thick solid curve). 
Then, we are lead to the conclusion that
the corona must have developed until $300 \Rg$ so as to have large 
values of $f$ and $\tc$ at the outermost ring, thereby reproducing the
observed spectrum. 

\subsection{Red Optical Spectrum}

Optical spectral index of QSOs has been recognized as a discrepancy
between disk models and observations; Francis et al. (1991) 
reported that the bright QSOs, in which stellar contamination to the
optical fluxes is relatively negligible, 
typically show spectral index in optical,
$\alpha$ at 1500--5000 ${\rm \AA}$, 
of 0.32 ($L_{\nu} \propto \nu^{- \alpha}$), while the classical
standard disk model predict that of $-0.33$ 
(i.e., $L_\nu \propto \nu^{1/3}$).

In order to emphasize the optical spectra, a part of Fig.\ 2a is
enlarged and shown in figure 3 with additional spectra of disk models
integrated over 3 $< R < 1000 R_g$.
Thick solid and dotted lines have the same meanings as in Fig.\ 2a, 
and they both show similar spectral index, $\alpha \sim 0.3$.
The shaded area inndicates the relevant waveband, 1500--5000 ${\rm \AA}$.

The reason why the current model has a better agreement with the
observations in terms of the optical spectral index is as follows.
The long-dashed line is
a spectrum of the standard disk (i.e., eq.\ 4), 
but without the inner boundary term (the term within the parenthesis on
the right-hand-side of eq.\ 4, 
which is sometimes omitted when roughly estimating the spectral index 
of the disk model).
It shows $\alpha \sim -0.3$.
Spectra of the standard disks with the inner boundary term are drawn by
the short-dashed lines.
The upper one represents the disk with $\Mbh = 3 \times 10^9 \Msun$ and
$\Mdot = 0.5 \, \Ledd / c^2$, while the lower one is for 
$\Mbh = 10^8 \Msun$ and $\Mdot = \Ledd / c^2$ showing 
somewhat redder spectrum ($\alpha\sim -0.1$) due to the inner-boundary term.

The emergent local spectrum changes its shape in a sense that 
the spectrum is distorted and shifted towards shorter wavelength 
when the electron 
scattering in the disk is taken into account (Czerny \& Elvis 1987).
Such a spectral shift 
is more efficient in inner region than
in outer region.
As the result, disk models with electron scattering finally show 
redder optical spectra, 
$\alpha \sim +0.1$ in lower solid curve and $\alpha \sim +0.3$ in 
upper solid curve.
The difference of spectral indices in the two solid curves, 
which are obtained by radial integrations of ST93 model (i.e., no
corona), are due mainly to the difference of their black-hole masses.
The
disk with larger $M_{\rm BH}$ gets cooler at fixed $\dot{M} / L_{\rm Edd}$
and $R/R_g$, as is expected for the standard accretion disk 
(eq.\ 14 described later).
The optical band becomes to measure the flux at inner part of the disk,
where it gets close to the edge of the integral interval (e.g., 3 $< R <
1000 R_g$).
Then, the emergent spectrum is somewhat curved in a sense that 
it is redder than
the spectrum from outer region (e.g., figure 16 in Koratkar \& Blaes 1999).
Note that the redder spectra are achieved even within the framework of 
original ST93 and ST95 models, as well as our disk-corona model.
If one wants to discuss in more details about optical spectra,
heavy metals, bound-bound, free-bound, bound-free transitions and 
precise estimation of the contribution from broad line region 
(e.g., Balmer continuum) must be included.

\subsection{Spectra for Various $\Mbh$ and $\Mdot$}

Figure 4 shows the black-hole mass dependence of the spectra with a
fixed ratio of $\Mdot$ to $\Ledd$ (i.e., fixed $\Mdot / \Mbh$).
The peak frequency of the big blue bump 
varies with $\Mbh$ in a similar fashion to 
that of the standard accretion disk in which
\begin{eqnarray}
T_{\rm eff} (4 \Rg) \propto 
 \Mbh^{-1/2} \Mdot^{1/4} 
 \ \propto \ \Mbh^{-1/4} (\Mdot / \Ledd)^{1/4}.
\end{eqnarray}
On the other hand, the cut-off frequency of the hard X-ray (i.e., 
coronal electron temperature) and the spectral slopes in X-rays 
are rather insensitive to the black-hole mass.

Accretion-rate dependence of the emergent spectra with a fixed blackhole
mass ($\Mbh = 3 \times 10^9 \Msun$) is shown in Figure 5.
The spectral slope at 0.1 -- 2.0 keV stays almost constant, 
$\alpha \!$ = 1.46, 1.58, 1.62
for $\Mdot / (L_{\rm Edd} / c^2) \!$ = 0.1, 0.5 and 0.7, respectively.

\subsection{Vertical Structure}

We finally present the vertical structure of the disk/corona.
Number density of electrons (dotted line) and temperature of protons
(crosses) and electrons (solid line) at $R \sim 4.9 \Rg$ are shown in Figure 6.
The left side of the figure where $\xi$ [$\equiv$ log($\t0 - \tau$)] = 3
corresponds to the mid-plane (i.e., $\t0 = 1000$), 
and the boundary between the disk and corona
is located at $\xi \sim - 0.2$ (i.e., $\tc$ = 0.6). 
Parameters used here are the same as those of Fig.\ 2a.
It turns out that the height of the boundary measured from the mid-plane is
0.03 $\Rg$, and that of the surface of the corona (at $\xi$ of $-2$) is
0.3 $\Rg$.
Then, the disk-corona system is indeed geometrically thin;
the height of the corona from the mid-plane $\lesssim \, 0.1 R$.

\section{DISCUSSION AND CONCLUSION}

\subsection{Comparison with previous models}

The most widely accepted interpretation, to date, 
of soft and hard X-ray emission
mechanisms is that soft excess is due to unscattered photons 
propagating from the disk-corona boundary to the coronal surface, 
while hard power-law emission is attributed to unsaturated
Comptonization, i.e., photons that are Compton up-scattered during the
propagation within the corona.
Monte-Carlo simulations of propagating photons in the framework of two zone
treatment (disk and coronal layers) indeed 
reproduce bump-like feature in soft 
X-ray and power-law component due to Comptonization in hard X-ray 
(e.g., Haardt \& Maraschi 1991; Nakamura \& Osaki 1993).
However, the bump-like feature due to unscattered photons 
still keeps its spectral shape as it was at the disk-corona 
boundary even after it goes away from the coronal surface: 
i.e., it should look like blackbody radiation.
Then, higher energy tail of the soft excess 
is predicted to be as steep as the Wien law,
in contradiction with observed small spectral indices in soft X-ray 
($\alpha \sim 1.4$ -- $1.6$).

Thermal bremsstrahlung emission was proposed for hard X-ray power-law
component by Schlosman, Shaham \& Shaviv (1984), in which the emission comes
from the disk-corona transition layer with an temperature gradient, 
where thermal conduction from the overlying corona is balanced with
radiative cooling.
They show that such a layer emits hard X-rays efficiently 
in the case of an accretion disk model proposed by
Sakimoto \& Coroniti (1981), 
in which viscous torque is proportional to gas pressure, 
but that it does not work in standard accretion disks 
where the torque is in proportion to the total pressure 
unless viscosity parameter in the disk is less than $10^{-3}$.
We should note here that non-thermal, power-law electron energy
distribution models, similarly to our results, 
produce two spectral power-laws for soft and hard X-ray emissions
(e.g., Zdziarski \& Lightman 1985),
although ours is a thermal model.

Shimura et al.\ (1995) proposed a model in which 
FUV--hard X-ray is attributed to unsaturated Comptonization in a corona
above a disk. 
Their claim is that their model, with $\tc$ being a strongly 
increasing function with increasing radius, 
reproduces a single power-law in X-ray bands, 
not a broken power-law (i.e., soft excess and hard power-law). 
Then, the corona could be patchy so that the disk is covered over only 
in part with the corona.
In our model, however, bremsstrahlung emission from the corona 
and reflection of the
coronal emission at the disk-corona boundary also contribute to 
hard X-ray as well as the unsaturated Comptonization. 
As a result, we have different spectral slopes at soft and hard X-ray bands.
It largely depends on our assumption of constant $f$ and $\tc$.
In other words, we need $f$ and $\tc$ to be almost constant 
or weak functions of radius
in order to expect sufficient bremsstrahlung emission from the corona
at outer region and then in order to 
reproduce the observed spectrum within the current framework.
Theoretical works to check the assumption are needed as next steps.
Without advective cooling in the corona, we also have rather straight
spectrum from FUV to hard X-ray (see Fig.\ 2c). 
The effect of advection seems to bent the total spectrum so as that 
the spectrum has two power-laws in X-ray bands.

In the extremely high accretion rate, the disk spectra may 
be able to reproduce the observed steep spectral decline 
at $1000\rmAA$--$400\rmAA$,
since Comptonization is more efficient at higher accretion rates
and since an increase of accretion rate changes the disk dynamics from the
standard disk to optically-thick ADAF.
Optically-thick ADAF, which is so called the slim disk model, 
radiates with higher temperature than the standard disk
(Szuszkiewicz, Malkan \&  Abramowicz 1996; Mineshige et al.\ 2000).
Narrow-Line Seyfert 1 galaxies (NLS1s) have peculiar spectral 
and temporal features that are not seen in normal Seyfert nuclei and QSOs 
(e.g., Boller, Brandt, \& Fink 1996; Brandt, Mathur, \& Elvis 1997; 
Leighly 1999a, 1999b; Grupe et al.\ 1998, 1999).
These features are often attributed to small black-hole mass and to high 
accretion rate (i.e., $L \sim \Ledd $; e.g., Pounds, Done, \& Osborne 1996; 
Hayashida et al.\ 1998; Mineshige et al.\ 2000).
Then, such extreme accretion rate is not relevant in this study, 
where we are going to
construct a disk model for normal Seyferts/QSOs that perhaps have 
moderate accretion rate [$L \sim$ (0.01 -- 0.1) of $\Ledd$; e.g., Wandel 1999].
To account for systematically large $\alpha_{ROSAT}$ in NLS1s,
we would need to reduce Compton-$y$ parameter in the corona.

The configuration of cold ($\sim$ 10$^5$ K) 
and hot ($\sim$ 100 keV) regions 
that radiate optical/UV bump and X-rays, respectively, 
is an unsolved problem. 
Optically-thin ADAF 
is a possible energy source of X-ray emissions. 
However, optically-thin ADAF itself is basically faint 
(e.g., Mahadevan 1997) compared with observed luminosity of AGNs. 
Then, the luminosity of the power-law X-ray emission 
indicates that cold material is located adjacent to such hot flow as a
source of seed photons: for example, 
corona-like flows above and below a disk
(e.g., Liang \& Price 1977; Haardt \& Maraschi 1991).
Separation in the vertical direction 
has been also suggested from 
the stability against thermal and secular instabilities in inner, 
radiation-pressure dominated region of the disk 
(Ionson \& Kuperus 1984; Nakamura \& Osaki 1993;
 Mineshige \& Kusunose 1993),
the correlated variabilities between the iron-line flux and the X-ray
continuum (e.g., Nandra \ea 1999; see also, however, Lee \ea 1999), 
and need for a cold disk as a `mirror' 
for reflection-component/broad-fluorescence line 
(e.g., Tanaka et al.\ 1995; Mushotzky \ea 1995; Nandra \ea 1997).
Another potential problem is that those 
layers could be inhomogeneous and highly time dependent;
the corona can be patchy (Shimura, Mineshige, \& Takahara 1995; 
Di Matteo 1998; Kawaguchi et al.\ 2000; Machida et al.\ 2000), 
or the cold gas may exist as numerous blobs within the corona instead 
of the slab geometry (Guilbert \& Rees 1988; Lightman \& White 1988;
Sivron \& Tsuruta 1993; Collin-Souffrin \ea 1996; 
Krolik 1998; R\'{o}\.{z}a\'{n}ska 1999). 
Nevertheless, spatially and temporally averaged treatment of the cold
and hot regions will still be useful for study of time-averaged spectrum. 
Appropriate treatment of mass evaporation/condensation process is
another future issue.
In other words, $f$ and $\tc$ may be variable in terms of 
$R$, $\Mbh$, $\Mdot$, etc. 
For instance, see Janiuk \ea (2000) for $f$ as a function
of $R$ with various $\Mdot$ and viscosity parameter, 
and \.{Z}ycki \ea (1995) for $\tc$ v.s.\ $\Mdot$.

\subsection{Expected Time Variability}

Time variability provides strong constraints on spectral models. 
Then, it will be worthwhile to briefly comment on the temporal 
behaviour, although we are now working on a model for the 
steady accretion disk-corona with an aim to reproduce the observed, 
time-averaged spectrum.
Followings are qualitative arguments expected from our current model.
More details are out of the scope of this paper, 
and will be remained as future works.

 For the optical-to-UV time-lag observed in NGC 7469 
(Wanders et al.\ 1997; Collier et al.\ 1998) and marginally detected 
in NGC 4151 (Peterson et al.\ 1998), our model can 
 explain the lag as Collier et al.\ (1998) demonstrated using the standard 
 accretion disk model.
That is because optical/UV emitter in our model is, 
like other spectral models, an
 accretion disk that has a similar dependence of (surface) temperature
 upon radius to the standard accretion disk.
It should be noted here that such a time-lag has not yet been established 
in all AGNs (Edelson et al.\ 2000)

Since the FUV, soft X-ray and a part of hard X-ray emission come from
the same region in our model, 
they are expected to vary without serious time-lag 
(larger than an order of days),
except for difference of escaping (diffusion) time at different energies.
Chiang et al.\ (2000) reported the inter-band lags in NGC 5548 
from $\sim$ 240 k\,sec simultaneous observations;
0.14-0.18 keV flux leads 0.78 keV flux by $\sim$ 10 k\,sec, and 
5.4 keV flux by $\sim$ 30 k\,sec. 
As discussed in section 6 in their
paper, these lags are consistent with a coronal size of 10 $R_g$ 
for a $10^8 M_{\odot}$ black-hole.
Incidentally, the blackhole mass of NGC 5548 is estimated to be about 
$10^8 M_{\odot}$ through the reverberation mapping method 
(Wandel, Peterson \& Malkan 1999; Kaspi et al.\ 2000).
However, the $\sim$ 4-day X-ray lag behind UV near the peak flux levels
and almost simultaneous variations near the minimum fluxes 
detected in NGC 7469 
(Nandra et al.\ 1998) is not solved by our model easily, as 
well as other models, to date, fail to do.

Finally, our model qualitatively explains following two issues about 
spectral variability in X-rays on short timescales (less than months) 
that are reported by numerous X-ray observations (\S 3.2). 
i) Soft X-ray flux is more variable than hard X-ray flux.
ii) Hard X-ray becomes softer when it gets brighter.

\subsection{Conclusions}
We study emission spectrum emerging from 
the vertical 
disk-corona structure composed of two-temperature plasma by solving 
hydrostatic equilibrium and radiative transfer self-consistently. 
The key question is what physical condition
exhibits a soft X-ray excess with a spectral index $\alpha$ 
($\Fnu \propto \nu^{- \alpha}$)
of 1.6 and a hard X-ray tail with $\alpha$ of 0.7 at the same time.  
In our model, a fraction $f$ of viscous heating is assumed to be
dissipated in a corona with a Thomson optical depth of $\tc$, where
advective cooling is also included, and a remaining fraction, $1-f$,
dissipates within a main body of the disk. 
Our model can nicely reproduce 
the observed composite spectrum of AGNs, which shows
soft X-ray excess 
with $\alpha$ of about 1.5 and hard tail extending to 
$\sim$ 50 keV with a different slope ($\alpha \sim 0.5$). 
Our model should be checked with individual objects 
in the future, though.

The broken power laws ($\alpha \sim$ 1.5 below 2keV and $\sim$ 0.5
above) are the results of different emission mechanisms: 
unsaturated Comptonization in soft X-rays
and a combination of the Comptonization, bremsstrahlung, 
and a reflection of the coronal radiation at the disk-corona 
boundary in hard X-rays.
Previous models, where soft X-ray excess is attributed to blackbody or
saturated Comptonization of the disk blackbody, 
tended to deal with limited wavebands separately, while 
we tried to fit the broad-band SED simultaneously.
That is the reason why we propose here emission
mechanisms, that are different from previous 
models, in soft and hard X-ray bands.
Also, our model differs from traditional models in terms of 
X-ray emitting region; 
hard X-ray emission causes from spatially spread region up
 to 300 Schwarzschild radii.
The emergent optical spectrum is redder 
($\alpha \sim 0.3$) than that of the standard disk 
($\alpha \sim -0.3$), being consistent with observations,
due to the different efficiencies of spectral distortion and shift of disk 
emission at different radii.
The cut-off frequency of the hard X-ray (reflecting 
the coronal electron temperature) and X-ray spectral slopes 
are insensitive to the black-hole mass, 
while the peak frequency of the big blue bump 
is sensitive to the mass.

\acknowledgements
We wish to thank Hitoshi Negoro for helpful discussions and 
encouragement, Makoto Kishimoto, Masaru Matsuoka, 
Juri Poutanen, Ken Ebisawa and Lev Titarchuk for useful comments.
We would also like to thank an anonymous referee for rewarding
suggestions.
T.K. appreciates the organizers of the 
Guillermo Haro Advanced Lectures on the starburst-AGN connection 
held in Mexico, 2000, for providing him the excellent
lectures and acknowledges useful discussions with the participants,
especially Dario Trevese and Agnieszka Janiuk.
This work was supported in part by the Research Fellowship of the Japan
Society for the Promotion of Science for Young Scientists (4616, TK), 
and by the Grants-in Aid of the
Ministry of Education, Science, Sports and Culture of Japan (10640228, SM).

\medskip 
\centerline{
\vbox{\epsfxsize=7.5cm\epsfbox{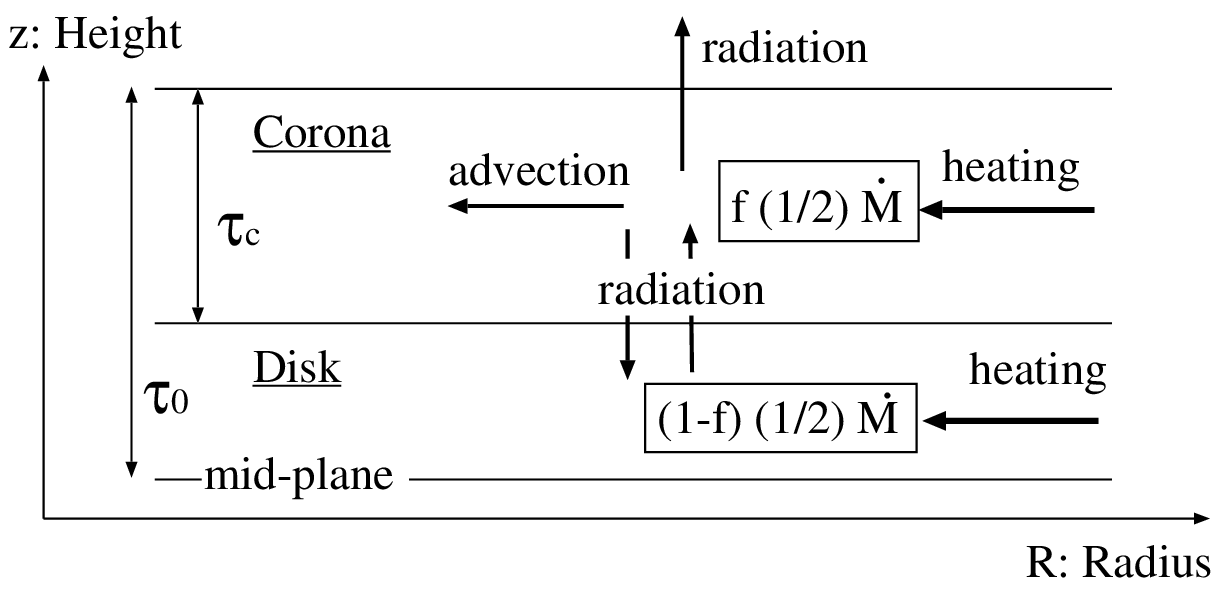}}}
{\small F{\scriptsize IG}.~1.--- 
A schematic view of the energy balance in the present model.}

\medskip 
\centerline{
\plotfiddle{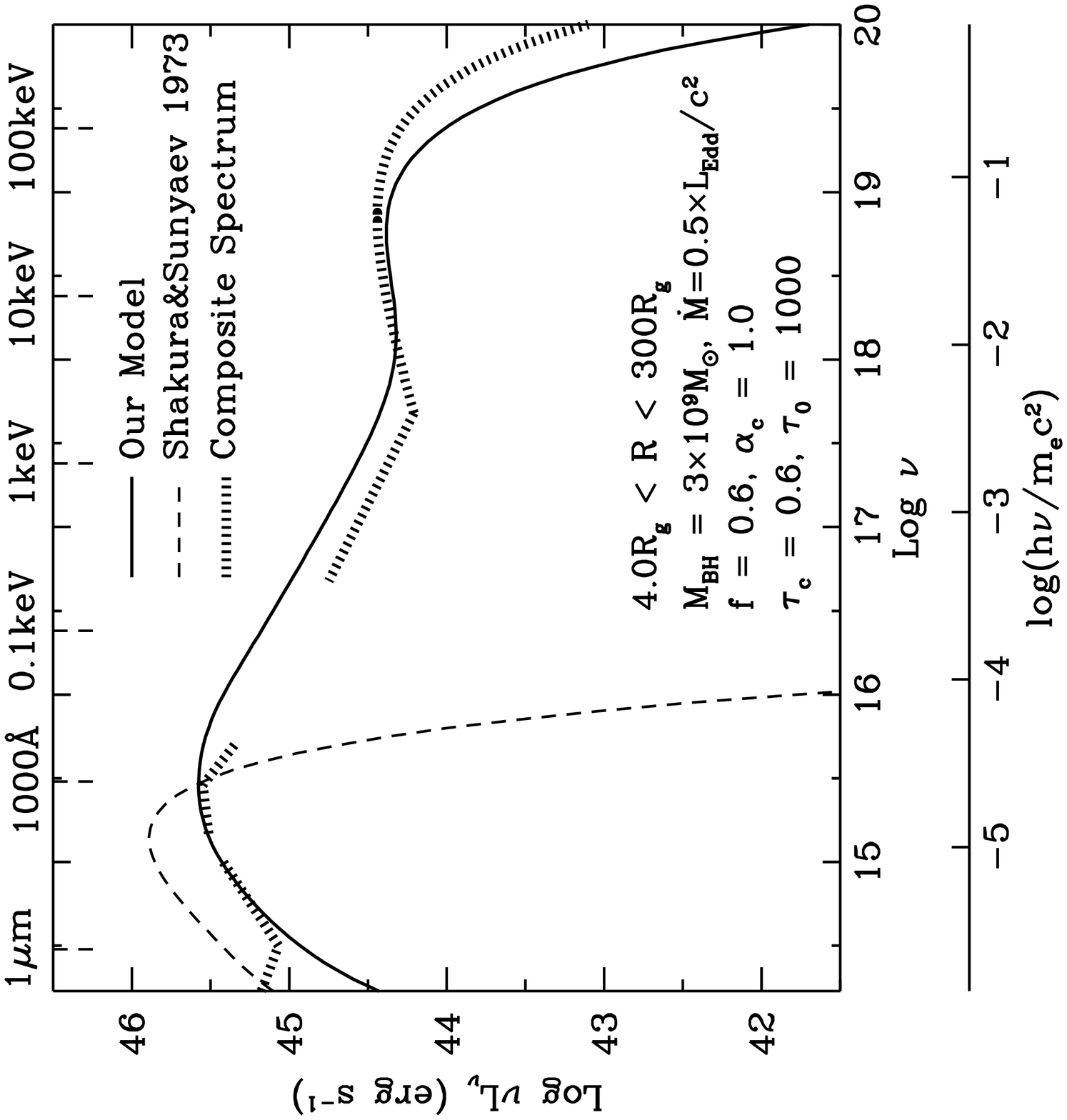}{100mm}{270}{60}{50}{-510}{300}}
{\small F{\scriptsize IG}.~2a.--- 
Resultant spectrum from the disk-corona structure 
integrated over 4.0 to 300$\Rg$ (thick line).
Parameters used in this model are listed in the figure.
With those parameter sets, advective cooling in the corona is comparable 
with the radiative cooling at $R \sim 4.9 \Rg$; 
$q^-_{\rm adv} / q^+_{\rm c} \sim 0.5$.
Dashed line indicates the integrated spectra of the standard disk 
with the same $\Mbh$ and $\Mdot$.
Spectral indices of the observed composite spectra (dotted lines) are: 
$\alpha$ of 1.4 at NIR ($\lambda \geq 1\mu m$), 
0.3 at optical ($1\mu m \geq \lambda \geq 2500 \rmAA$), 
1.0 at UV ($2500 \rmAA \geq \lambda \geq 1000 \rmAA$), 
1.8 at FUV ($1000 \rmAA \geq \lambda$), 
1.6 at soft X-ray (0.2--2.0 keV),
0.7 at hard X-ray ($>$ 2.0keV),
1.5 between optical and X-ray (2500 $\rmAA$--2keV). 
The energy cut-off for the hard X-ray power-law is assumed to be 100 keV.}

\medskip 
\centerline{
\plotfiddle{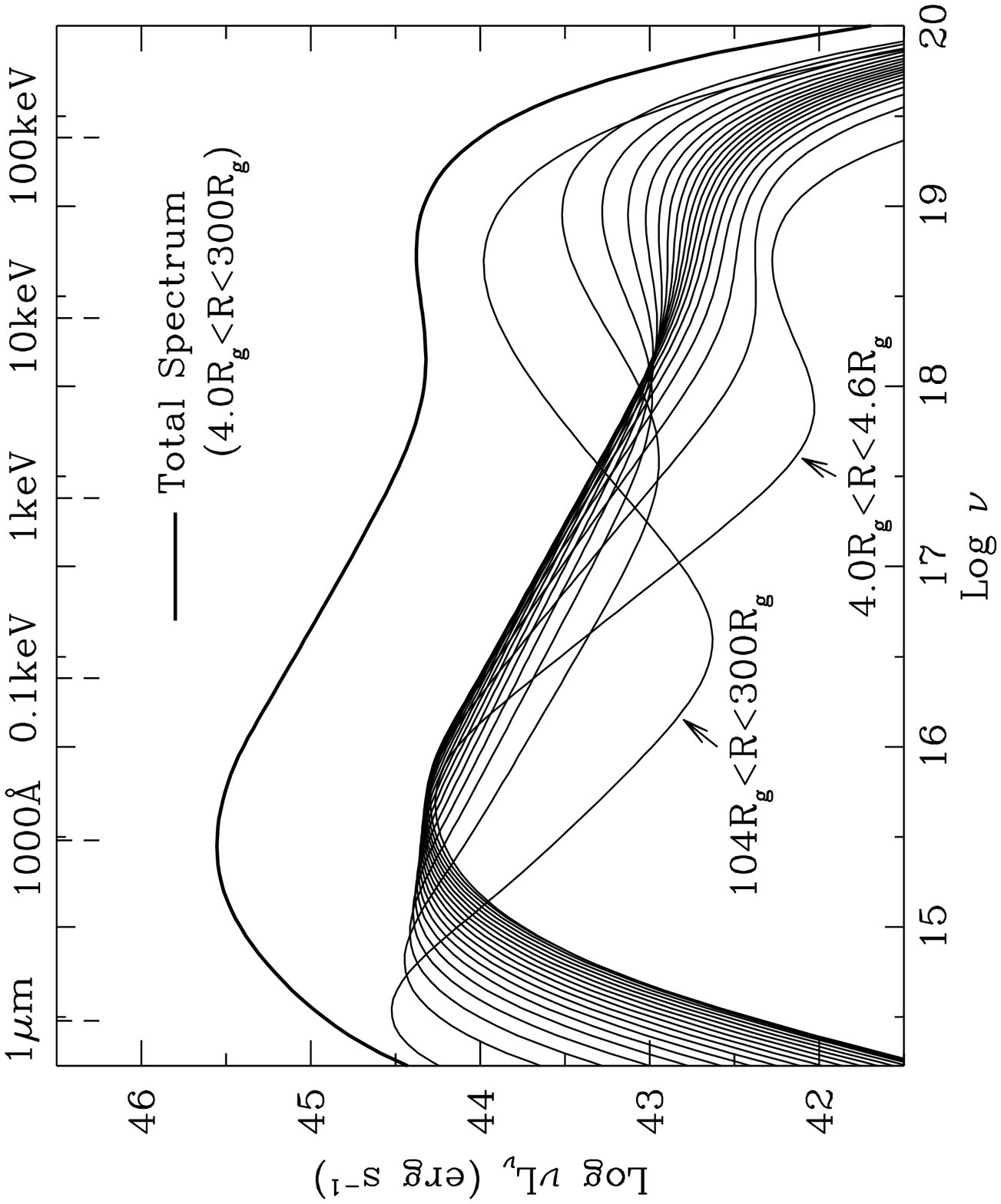}{100mm}{270}{60}{50}{-510}{265}}
{\small F{\scriptsize IG}.~2b.--- 
Contributions to the total spectrum due to individual rings (lower curves).
Thick line denotes the total spectrum ($4.0 \Rg < R < 300 \Rg$) 
as the same as in Fig.~2a.
}

\bigskip 
\centerline{
\plotfiddle{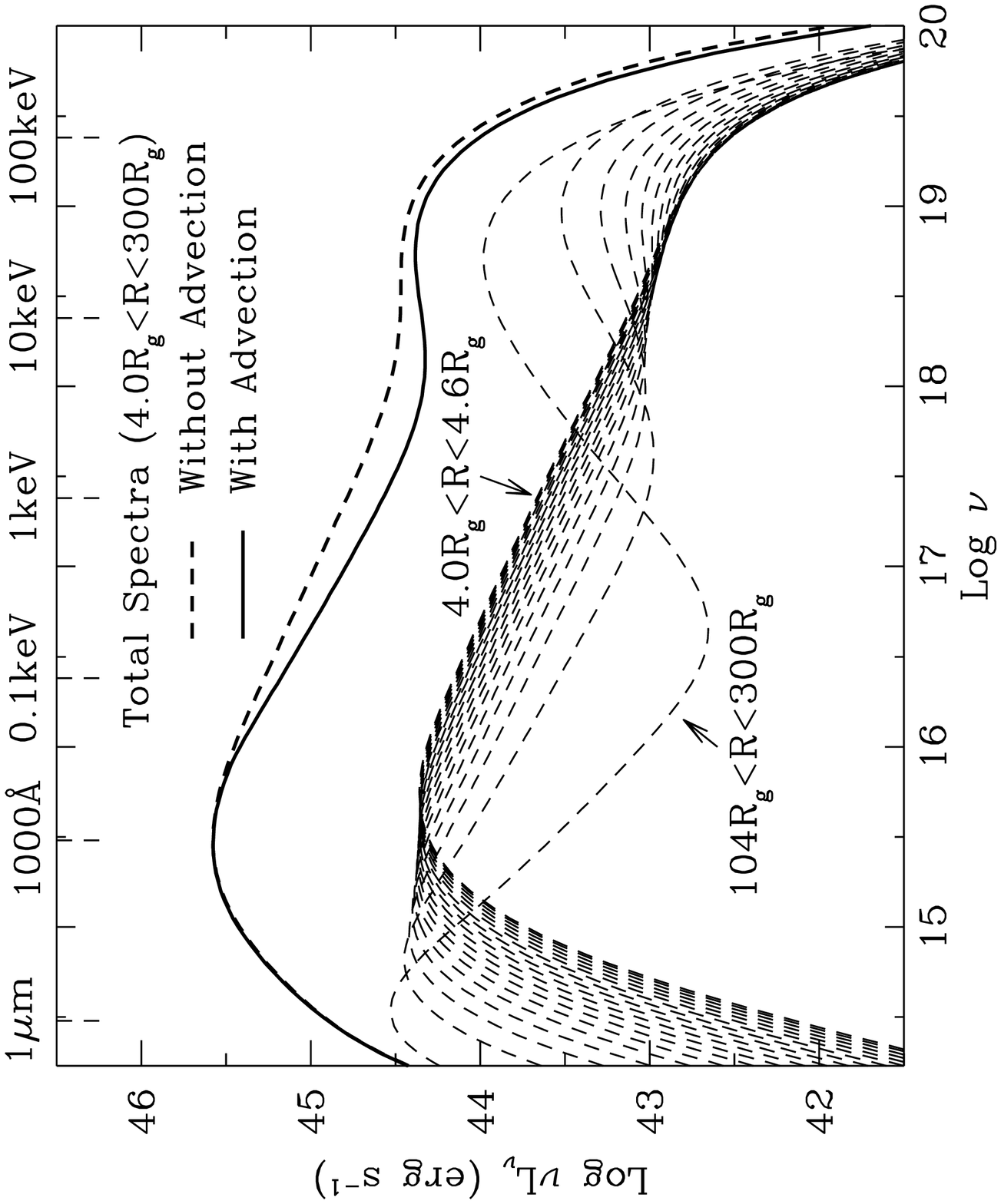}{100mm}{270}{60}{50}{-510}{265}}
{\small F{\scriptsize IG}.~2c.--- 
The effect of advective cooling in the
corona layer to the emergent spectrum.
Thick dashed line is the resultant spectrum without advective cooling.
Lower dashed curves represents contributions to the total spectrum 
due to each ring.
Thick solid curve has the same meaning as in Fig.\ 2a and 2b.}
\medskip

\medskip 
\centerline{
\plotfiddle{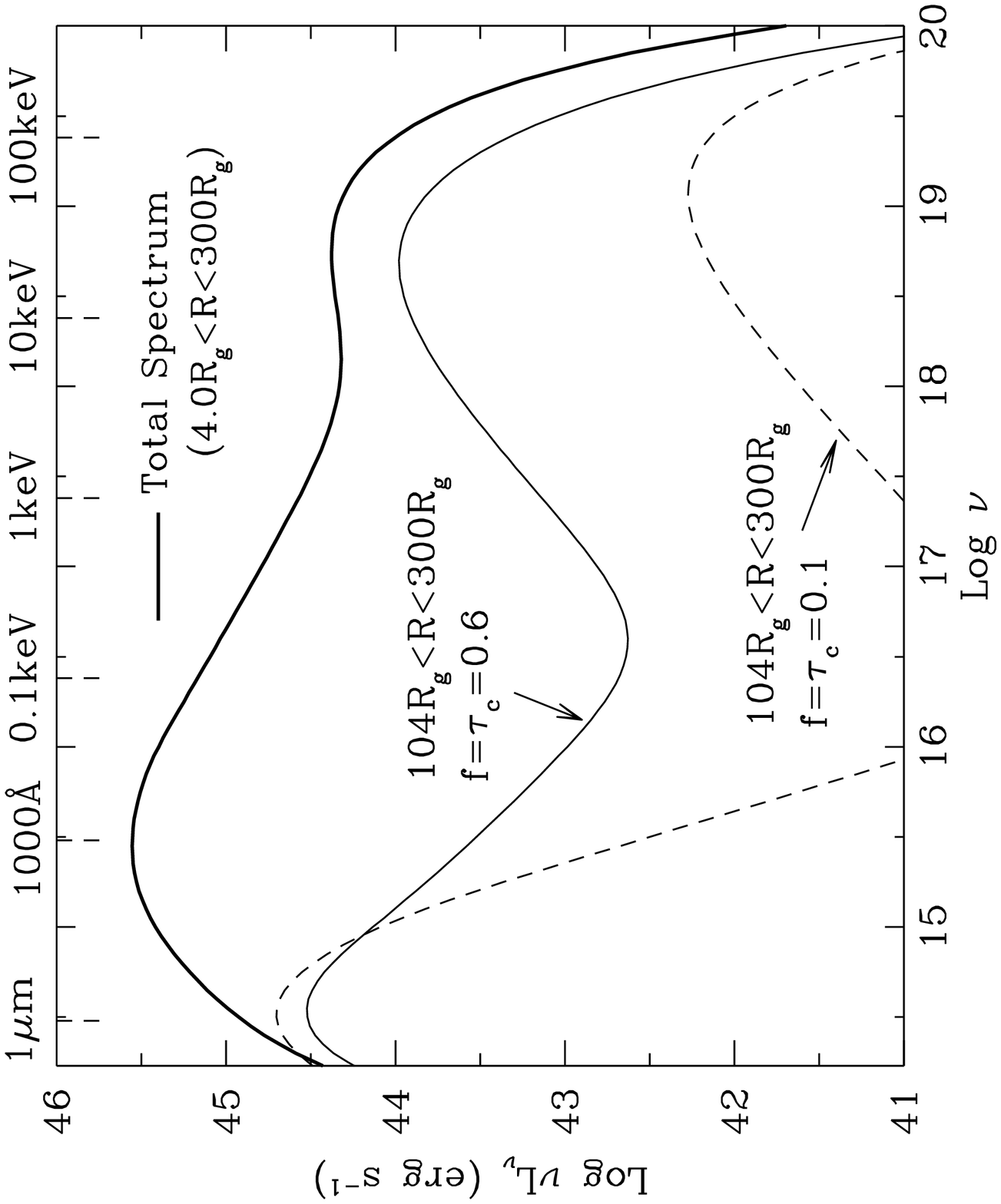}{100mm}{270}{60}{50}{-510}{265}}
{\small F{\scriptsize IG}.~2d.--- 
Spectra for the outermost ring with $f = \tc = 0.6$ (solid curve) and 
$f = \tc = 0.1$ (dashed curve).
It is shown that hard X-ray emission in the latter case
is not strong as the former
case, where bremsstrahlung emission at the outermost ring is the main
origin of hardening in the total X-ray spectrum (thick solid curve; see
Fig.\ 2b). 
}

\medskip 
\centerline{
\plotfiddle{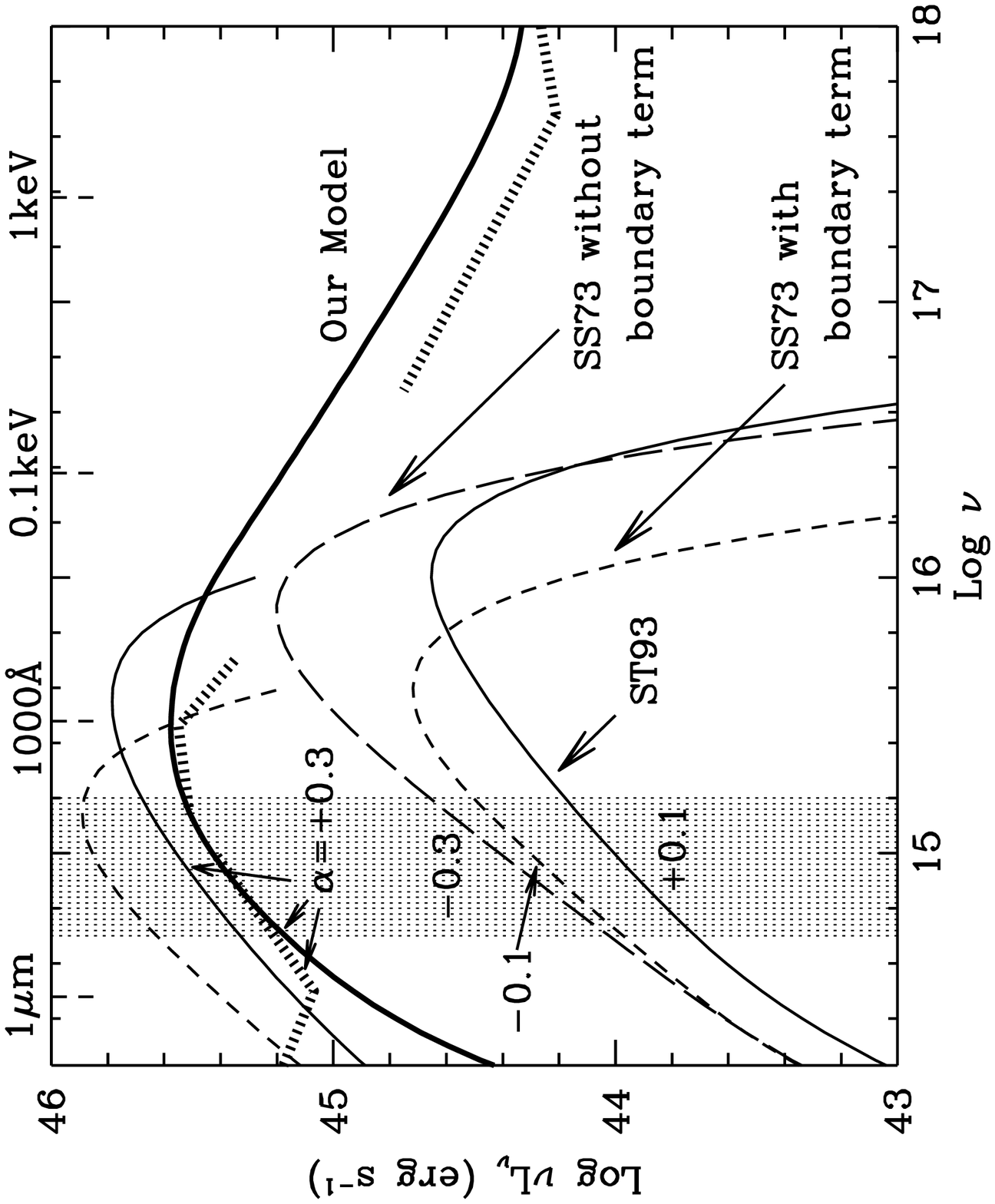}{100mm}{270}{60}{50}{-510}{265}}
{\small F{\scriptsize IG}.~3.---
Spectra of various models with optical spectral indices $\alpha$ 
($L_{\nu} \propto \nu^{- \alpha}$).
The shaded area inndicates the relevant waveband, 1500--5000 ${\rm \AA}$.
Thick solid and dotted lines have the same meanings as in Fig.\ 2a.
Spectra of the standard disks [Shakura \& Sunyaev 1973 (SS73), 
i.e., eq.\ 4] are shown as the
short-dashed lines as in Fig.\ 2a.
A long-dashed one is calculated through eq.\ 4 without inner boundary
term (the last term in eq.\ 4).
Finally, thin solid curves are obtained by radial integrations of ST93
model (i.e., no corona).
Lower three calculations are made for $M_{\rm BH} = 10^8 M_\odot$
and $\dot{M} = L_{\rm Edd}/c^2$ at 3 $< R < 1000 R_g$, 
while upper two curves are for $M_{\rm BH} = 3 \, 10^9 M_\odot$
and $\dot{M} = 0.5 \, L_{\rm Edd}/c^2$ at the same radii.
}

\medskip 
\centerline{
\plotfiddle{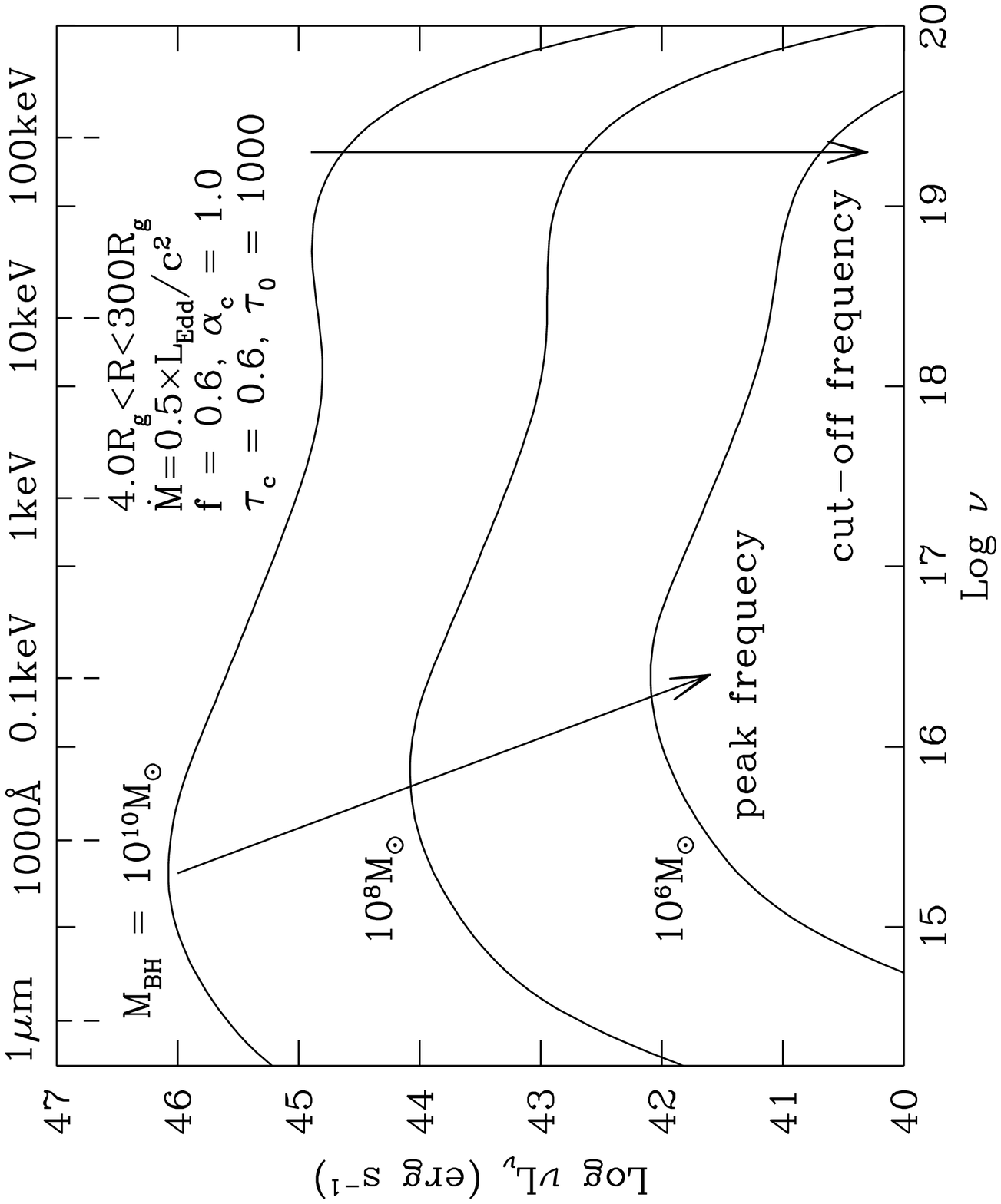}{100mm}{270}{60}{50}{-510}{265}}
{\small F{\scriptsize IG}.~4.--- 
Black-hole mass dependence of the emergent spectra 
with a fixed ratio of $\Mdot$ to $\Ledd$. 
Parameters used here are the same as those of Fig.\ 2a and 2b.
The cut-off frequency of the hard X-ray (reflecting 
the coronal electron temperature) and X-ray spectral slopes
are insensitive to the black-hole mass, 
while the peak frequency of the big blue bump is sensitive to the mass.} 

\medskip 
\centerline{
\plotfiddle{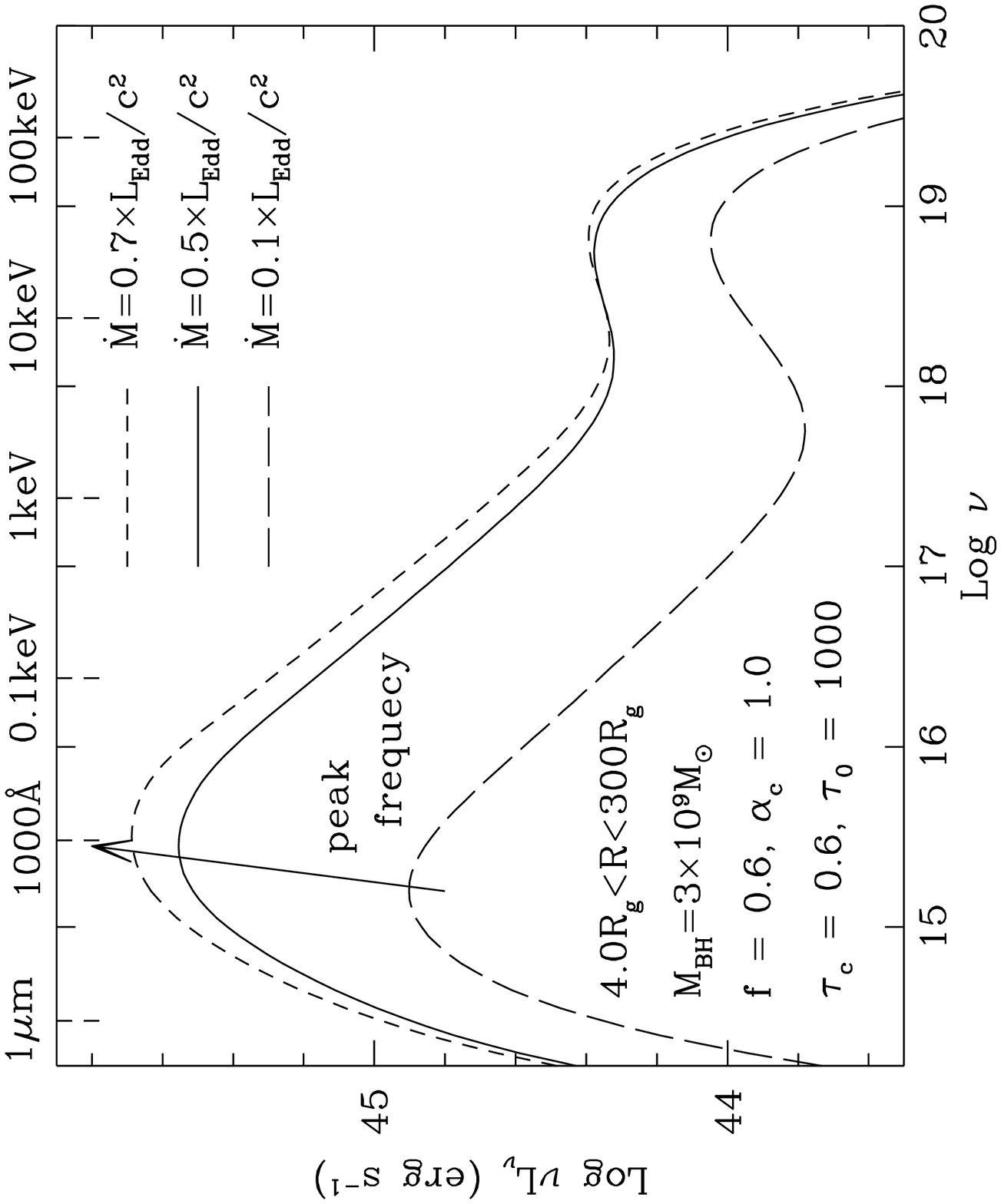}{100mm}{270}{60}{50}{-510}{265}}
{\small F{\scriptsize IG}.~5.--- 
Accretion-rate dependence of the emergent spectra with a fixed black-hole
mass ($\Mbh = 3 \times 10^9 \Msun$).} 

\medskip 
\centerline{
\plotfiddle{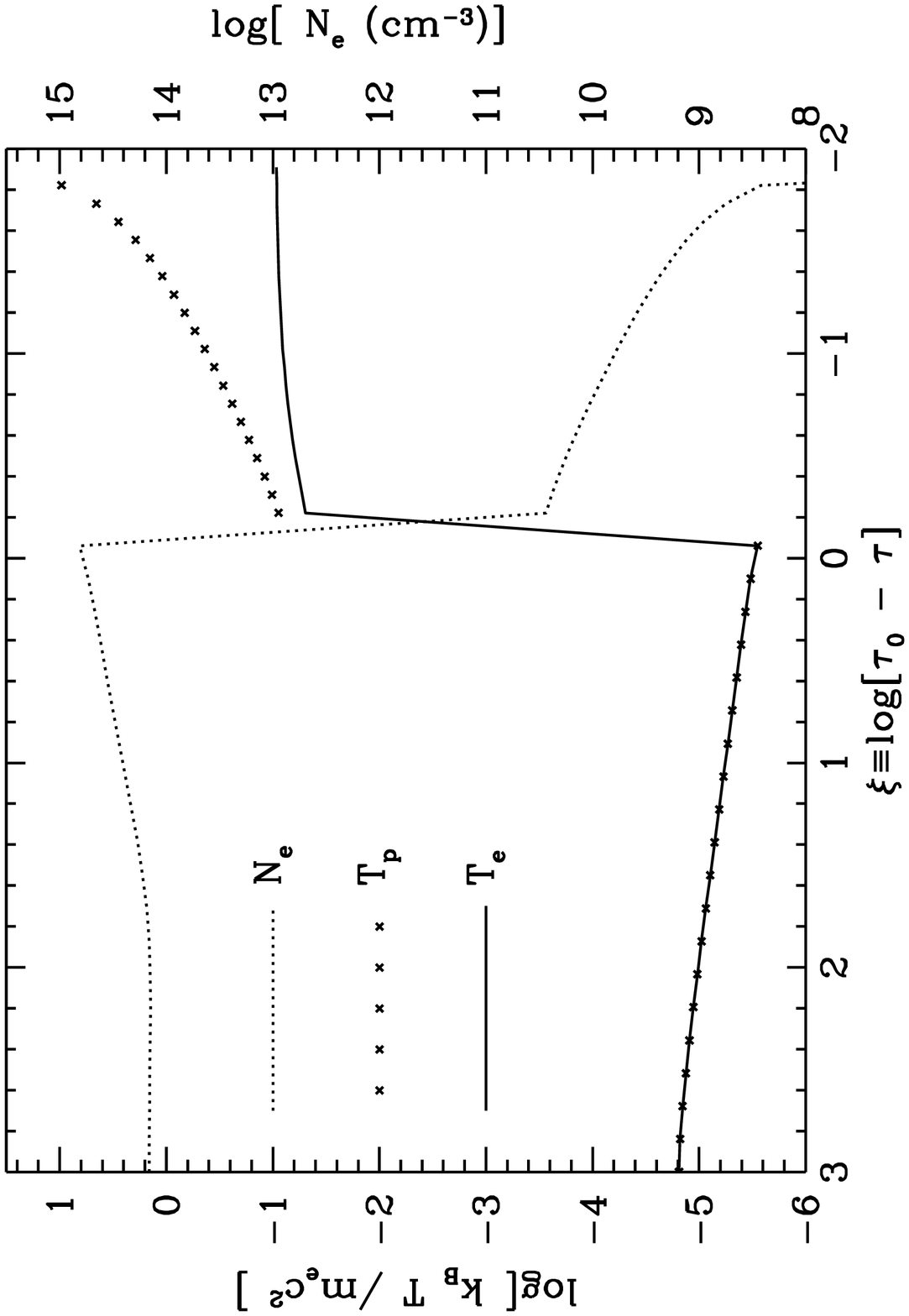}{100mm}{270}{60}{50}{-510}{265}}
{\small F{\scriptsize IG}.~6.--- 
Vertical structure of the matter density (dotted line), 
proton temperature (crosses), 
and electron temperature (solid line).
The mid-plane of the disk is located at $\xi$ of 3, 
and the boundary between the disk 
and corona is at $\xi \sim - 0.2$.
It turns out that the height of the boundary measured from 
the mid-plane is 0.03 $\Rg$, and that of the surface of the 
corona (at $\xi$ of $-2$) is 0.3 $\Rg$.
Note that $k_{\rm B}T$/$m_{\rm e} c^2$ = 1 corresponds to 
$T \simeq 5 \ 10^9$ K ($\simeq$ 500 keV).} 

\end{document}